\documentclass[fleqn,12pt,twoside]{article}
\usepackage{espcrc1}

\usepackage{graphicx}

\newcommand{\AmS}{{\protect\the\textfont2
A\kern-.1667em\lower.5ex\hbox{M}\kern-.125emS}}
\usepackage{amssymb}

\title{On the number of droplets in aerosols}

\author{H. Gouin\address {Universit{\'e} d'Aix-Marseille,  Lab.  Mod{\'e}lisation en M{\'e}canique et
Thermodynamique, \\ Ave Escadrille Normandie-Niemen, 13397 Marseille Cedex 20, France}%
        \thanks{e-mail: henri.gouin@univ-cezanne.fr},
        F. Cubisol\addressmark}

\begin{document}

\maketitle

\begin{abstract}
The number of droplets which may be formed with a supersaturated
vapor in presence of a gas cannot exceed a number proportional to
$(p_v-p_{v_0})^4$ where $p_v$ and $p_{v_0}$ denote  at the same
temperature the pressure of the supersaturated vapor-gas mixture and
the pressure of the saturated vapor-gas mixture. The energy
necessary to the droplet formation is also bounded by a number
proportional to $(p_v-p_{v_0})^2$.
\end{abstract}
\\
\\
Keywords: aerosols, droplets, liquid-vapor mixture.

\section{ INTRODUCTION}

Aerosols are usual in industries of cosmetics, pharmacy and food.
One also finds them in problems involving environment and pollution.
Epidemics, radioactive elements can be carried by small droplets;
volcanoes generate sulphated aerosols contributing to the
destruction of the ozone layer. The planetary atmospheres are
charged with such mixtures in suspension. The interaction between
oceans and atmosphere generates mixtures of gas and water vapor
partially condensed in form of fine droplets. Aerosols are also
created by sprays and are important even in the car industry on the
level of combustion in petrol and air mixture. The literature
abounds on the study of nucleation times, on the growth of germs in
vapor mixtures and on the motion of
microscopic droplets \cite{Djikaev,Levdansky}.\\
Aerosols consist of liquid droplets of very small radius (from 0.1
to 10 microns) in suspension in a vapor-gas mixture; the droplet
number is assumed weak enough to avoid  interactions (smaller than
one hundred thousand droplets per a cubic centimeter which
corresponds to a
volume proportion smaller than one for thousand).\\
In this paper, we prove that for given conditions of pressure and
temperature the number of droplets per unit volume cannot exceed a
maximum value. The value depends on the surface tension of
gas-vapor/liquid interface and on the difference between the
pressure of gas supersaturated with vapor and that of gas with
saturated vapor. The energy necessary to the creation of the
droplets is limited; providing to a gas-vapor mixture an energy
higher than this limit does not increase the droplet density.

\section{ POSITION OF THE PROBLEM}

In a large-sized isothermal tank of volume $V$, we consider a
mixture constituted of two fluids labelled with indices 1 and 2.
Fluid 1 is supposed at a temperature much lower than its critical
temperature and coexists in form of liquid and vapor bulks. Fluid 2
is supposed at a temperature much upper than its critical
temperature and behaves like a perfect gas (we call it the  gas). An
example of mixture verifying such conditions is that of water in
liquid and vapor forms in presence of air at the ambient
temperature. We propose to determine the equilibrium of $n$ liquid
droplets with a dissolved gas in  presence of supersaturated vapor
in the gas. The droplets form a fog and their volume fraction is
small enough so that they do not interact. The tank is large enough
so that the boundary effects are neglected. Let us denote
$\rho_{im}V$ the total mass of fluid $i\, (i=1,2)$. The mixture is
distributed between a gas-vapor bulk $v$ of volume $v_v$ and whose
densities are denoted $\rho_{iv}$ and the $n$ droplets of liquid
with dissolved gas of densities $\rho_{il} $ represented by a bulk
$l$. The radius droplet is small (of micron order) and the forces
due to gravity are negligible. At equilibrium, the droplets are
spherical and it is easy to notice that they are all identical with
a volume $v_l$. The volume free energy of the mixture is denoted
$\Psi(\rho_1,\rho_2)$. In a van der Waals model, it is a non convex
function of the densities $ \rho_1$ and $\rho_2$ \cite{gouin}. The
surface tension $\sigma$ of interfaces separating the bulks $l$ and
$v$ is assumed to be independent of $ \rho_1$ and $\rho_2$. This
approximation corresponds to the fact that density values will be
limited to small domains near the phase equilibrium. By taking into
account   the surface energy of the $n$ droplets, the total free
energy of the mixture in the tank is
\begin{equation}
E=n\Psi (\rho _{1l},\rho _{2l})v_{l}+\Psi (\rho _{1v},\rho _{2v})v_{v}+\frac{%
3}{2}\frac{n}{K^{\frac{1}{3}}}\, v_{l}^{\frac{2}{3}}
\end{equation}
with $\displaystyle K=\frac{3}{32\pi \sigma ^{3}}$. Volumes and
densities of the mixture bulks are submitted to the constraints
\begin{equation}
\left\{
\begin{array}{l}
nv_{l}+v_{v}=V \\
n\rho _{1l}v_{l}+\rho _{1v}v_{v}=\rho _{1m}V \\
n\rho _{2l}v_{l}+\rho _{2v}v_{v}=\rho _{2m}V
\end{array}
\right.
\end{equation}
The research of  mixture equilibrium reduces to study the extremum
of
\begin{equation}
F=E-\lambda (nv_{l}+v_{v})-\mu (n\rho _{1l}v_{l}+\rho
_{1v}v_{v})-\nu (n\rho _{2l}v_{l}+\rho _{2v}v_{v})
\end{equation}
Where $\lambda, \mu, \nu$ are three scalar Lagrange multipliers.
The balance equations associated with the variations of $v_{l},
v_{v}, \rho _{1l}, \rho _{2l}, \rho _{1v}, \rho _{2v}$ are
\begin{equation}
\left\{
\begin{array}{lll}
n\Psi (\rho _{1l},\rho _{2l})+ n (Kv_{l})^{-\frac{1}{3}} -n\lambda
-n\mu
\rho _{1l}-n\nu \rho _{2l} & = & 0 \\
\Psi (\rho _{1v},\rho _{2v})-\lambda -\mu \rho _{1v}-\nu \rho
_{2v} & = & 0
\\
n\Psi _{\rho _{1}}^{\prime }(\rho _{1l},\rho _{2l})v_{l}-\mu
nv_{l} & = & 0
\\
n\Psi _{\rho _{2}}^{\prime }(\rho _{1l},\rho _{2l})v_{l}-\nu
nv_{l} & = & 0
\\
\Psi _{\rho _{1}}^{\prime }(\rho _{1v},\rho _{2v})v_{v}-\mu v_{v} & = & 0 \\
\Psi _{\rho _{2}}^{\prime }(\rho _{1v},\rho _{2v})v_{v}-\nu v_{v}
& = & 0
\end{array}
\right.
\end{equation}
For the sake of simplicity, we denote also
\[\psi (l)=\Psi (\rho
_{1l},\rho _{2l}),\, \psi (v)=\Psi (\rho _{1v},\rho _{2v}), \,
\psi _{i}^{\prime }(l)=\Psi _{\rho _{i}}^{\prime }(\rho _{1l},\rho
_{2l}),\, \psi _{i}^{\prime }(v)=\Psi _{\rho _{i}}^{\prime }(\rho
_{1v},\rho _{2v})\,\] with  $\, i=1, 2.\,$  One deduces the
relations of equilibrium
\begin{equation}
\psi _{1}^{\prime }(l)=\psi _{1}^{\prime }(v)=\mu
\end{equation}
\begin{equation}
\psi _{2}^{\prime }(l)=\psi _{2}^{\prime }(v)=\nu
\end{equation}
\begin{equation}
\psi (v)-\rho _{1v}\psi _{1}^{\prime }(v)-\rho _{2v}\psi
_{2}^{\prime }(v)-\psi(l)+\rho _{1l}\psi _{1}^{\prime }(l)+\rho
_{2l}\psi _{2}^{\prime }(l)=(\frac{1}{Kv_{l}})^{\frac{1}{3}}
\end{equation}
and  the lagrange multiplier $ \lambda =\psi (v)-\rho _{1v}\psi
_{1}^{\prime }(v)-\rho _{2v}\psi _{2}^{\prime }(v).\, $ Relations
(5) and (6) express the equality of the chemical potentials of each
mixture component in the two bulks $l$ and $v$ (see Fig. 1). At the
temperature $T_0$, the total pressure value is
\[
p(\rho _{1},\rho _{2})=\rho _{1}\Psi _{\rho _{1}}^{\prime }(\rho
_{1},\rho _{2})+\rho _{2}\Psi _{\rho _{2}}^{\prime }(\rho
_{1},\rho _{2})-\Psi (\rho _{1},\rho _{2})+p_{o}
\]
where $p_0$ is a reference pressure and relation (7) is equivalent
to
\begin{equation}
p_l-p_v=\frac{2\sigma }{R}
\end{equation}
where $R$ is the radius of the droplets and $p_l, p_v$ denote the
pressure in  bulks $l$ and $v$. Relation (8) is the expression of
Laplace equation for spherical interfaces of fluid mixtures
\cite{Rocard}.

\begin{figure}[h]
\begin{center}
\includegraphics[width=9cm]{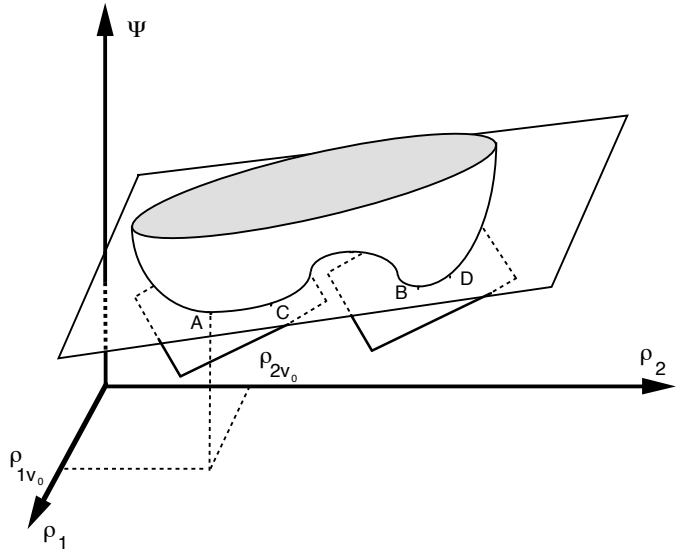}
\end{center}
\caption{For a plane interface, the balance of phases is
associated with contact points  A and B of a bi-tangential plan to
the free energy surface. Points C and D correspond to the balance
of phases between a phase corresponding to droplets and the
gas-vapor bulk. Tangent plans at points C and D to the free energy
surface are parallel \cite{Rowlinson}.} \label{fig1}
\end{figure}

\section{ THE DROPLET NUMBER PER UNIT VOLUME OF THE MIXTURE}

Relations (2)  make it possible to write:
\begin{equation}
 v_{l}=\frac{\rho _{im}-\rho _{iv}}{n(\rho _{il}-\rho
_{iv})}\ V \qquad \qquad v_{v}=\frac{\rho _{il}-\rho _{im}}{\rho
_{il}-\rho _{iv}} \ V\qquad \qquad(i=1,2)
\end{equation}
Expression (7) yields
\begin{equation}
N=K\frac{\rho _{im}-\rho _{iv}}{\rho _{il}-\rho _{iv}}\left[
p(\rho _{1l},\rho _{2l})-p(\rho _{1v},\rho _{2v})\right] ^{3}
\end{equation}
where $N = n/V$ is the number of droplets per unit volume of the
mixture. Let us notice that $\displaystyle 0\leq \frac{\rho
_{im}-\rho _{iv}}{\rho _{il}-\rho _{iv} }\leq 1$. For $\rho
_{iv}=\rho _{iv_{o}}\, \mathrm{then}\,\, \rho _{il}=\rho
_{il_{o}}$ where $\rho _{iv_{o}}, \rho _{il_{o}}$ are the
densities in the bulks of component $i$ for a liquid-vapor plane
interface, we obtain $p(\rho _{1l_{o}},\rho _{2l_{o}})=p(\rho
_{1v_{o}},\rho _{2v_{o}})$ and $N = 0$.  In the same way when
$\rho _{iv}=\rho _{im}, \, $ then $ \, N = 0$.  Consequently, the
number of drops per unit volume admits a maximum when $\rho_{iv}
\in \left] \rho _{iv_{o}},\rho _{im}\right[$. In the  case for
which $\rho _{1m}$ and $\rho _{2m}$ are close from $\rho
_{1v_{o}},\rho _{2v_{o}}$, explicit calculations are simplified.

\noindent Owing to the fact we can chose the chemical potentials
and the free energy as
\[
 \Psi (\rho _{1l_{o}},\rho _{2l_{o}})=\Psi (\rho
_{1v_{o}},\rho _{2v_{o}})=0
\]
\[
\Psi _{\rho _{1}}^{\prime }(\rho _{1l_{o}},\rho _{2l_{o}})=\Psi
_{\rho _{1}}^{\prime }(\rho _{1v_{o}},\rho _{2v_{o}})=0
\]
\[
\Psi _{\rho _{2}}^{\prime }(\rho _{1l_{o}},\rho _{2l_{o}})=\Psi
_{\rho _{2}}^{\prime }(\rho _{1v_{o}},\rho _{2v_{o}})=0,
\]
near  equilibrium positions,    the free energy expansions are
\[
\begin{array}{l}
\Psi (\rho _{1l},\rho _{2l})\;=\;\;\frac{1}{2}(\rho _{1l}-\rho
_{1l_{o}})^{2}\Psi _{\rho _{1}^{2}}^{^{\prime \prime }}(\rho
_{1l_{o}},\rho
_{2l_{o}}) \\
\;\;+(\rho _{1l}-\rho _{1l_{o}})(\rho _{2l}-\rho _{2l_{o}})\Psi
_{\rho
_{1}\rho _{2}}^{^{\prime \prime }}(\rho _{1l_{o}},\rho _{2l_{o}})+\frac{1}{2}%
(\rho _{2l}-\rho _{2l_{o}})^{2}\Psi _{\rho _{2}^{2}}^{^{\prime
\prime
}}(\rho _{1l_{o}},\rho _{2l_{o}}) \\
\Psi (\rho _{1v},\rho _{2v})\;=\;\;\frac{1}{2}(\rho _{1v}-\rho
_{1v_{o}})^{2}\Psi _{\rho _{1}^{2}}^{^{\prime \prime }}(\rho
_{1v_{o}},\rho
_{2v_{o}}) \\
\;\;+(\rho _{1v}-\rho _{1v_{o}})(\rho _{2v}-\rho _{2v_{o}})\Psi
_{\rho _{1}\rho _{2}}^{^{\prime \prime }}(\rho _{1v_{o}},\rho
_{2v_{o}})+\frac{1}{2} (\rho _{2v}-\rho _{2v_{o}})^{2}\Psi _{\rho
_{2}^{2}}^{^{\prime \prime }}(\rho _{1v_{o}},\rho _{2v_{o}})
\end{array}
\]
and the chemical potential expansions are
\begin{equation}
\begin{array}{l}
\Psi _{\rho _{1}}^{\prime }(\rho _{1v},\rho _{2v})=(\rho
_{1v}-\rho _{1v_{o}})\Psi _{\rho _{1}^{2}}^{^{\prime \prime
}}(\rho _{1v_{o}},\rho _{2v_{o}})+(\rho _{2v}-\rho _{2v_{o}})\Psi
_{\rho _{1}\rho _{2}}^{^{\prime
\prime }}(\rho _{1v_{o}},\rho _{2v_{o}}) \\
\Psi _{\rho _{1}}^{\prime }(\rho _{1l},\rho _{2l})\;=(\rho
_{1l}-\rho _{1l_{o}})\Psi _{\rho _{1}^{2}}^{^{\prime \prime
}}(\rho _{1l_{o}},\rho _{2l_{o}})+(\rho _{2l}-\rho _{2l_{o}})\Psi
_{\rho _{1}\rho _{2}}^{^{\prime
\prime }}(\rho _{1l_{o}},\rho _{2l_{o}}) \\
\Psi _{\rho _{2}}^{\prime }(\rho _{1v},\rho _{2v})=(\rho
_{1v}-\rho _{1v_{o}})\Psi _{\rho _{1}\rho _{2}}^{^{\prime \prime
}}(\rho _{1v_{o}},\rho _{2v_{o}})+(\rho _{2v}-\rho _{2v_{o}})\Psi
_{\rho _{2}^{2}}^{^{\prime \prime
}}(\rho _{1v_{o}},\rho _{2v_{o}}) \\
\Psi _{\rho _{2}}^{\prime }(\rho _{1l},\rho _{2l})\;=(\rho
_{1l}-\rho _{1l_{o}})\Psi _{\rho _{1}\rho _{2}}^{^{\prime \prime
}}(\rho _{1l_{o}},\rho _{2l_{o}})+(\rho _{2l}-\rho _{2l_{o}})\Psi
_{\rho _{2}^{2}}^{^{\prime \prime }}(\rho _{1l_{o}},\rho
_{2l_{o}})
\end{array}
\end{equation}
From the equality of chemical potentials in the two bulks $l$ and
$v$, one obtains
\[2\left(p_l-p_v\right)=(\rho _{1l}+\rho
_{1l_{o}}-\rho _{1v}-\rho _{1v_{o}}) \psi^\prime_1(v)+ (\rho
_{2l}+\rho _{2l_{o}}-\rho _{2v}-\rho _{2v_{o}})\psi^\prime_2(v)
\]
and from relations (11),
\[
\begin{array}{l}
2\left(p_l-p_v\right)\,=\quad (\rho _{1v}-\rho _{1v_{o}})(\rho
_{1l}+\rho _{1l_{o}}-\rho _{1v}-\rho _{1v_{o}})\Psi
_{\rho _{1}^{2}}^{^{\prime \prime }}(\rho_{1v_{o}},\rho_{2v_{o}})\\
\quad \quad \quad \quad \quad \quad +\, \,(\rho _{2v}-\rho
_{2v_{o}})(\rho _{1l}+\rho
_{1l_{o}}-\rho _{1v}-\rho _{1v_{o}})\Psi _{\rho _{1}\rho _{2}}^{^{\prime \prime }}(\rho_{1v_{o}},\rho_{2v_{o}})\\
\quad \quad \quad \quad \quad \quad +\, \,(\rho _{1v}-\rho
_{1v_{o}})(\rho _{2l}+\rho
_{2l_{o}}-\rho _{2v}-\rho _{2v_{o}})\Psi _{\rho _{1}\rho _{2}}^{^{\prime \prime }}(\rho_{1v_{o}},\rho_{2v_{o}})\\
\quad \quad \quad \quad \quad \quad +\, \,(\rho _{2v}-\rho
_{2v_{o}})(\rho _{2l}+\rho _{2l_{o}}-\rho _{2v}-\rho
_{2v_{o}})\Psi _{\rho _{2}^{2}}^{^{\prime \prime }}(\rho_{1v_{o}},\rho_{2v_{o}})
\end{array}
\]
Vapor and  gas constitute a perfect gas mixture and the volume
free energy is the sum of the volume free energies of each fluid
In bulk $v$,
\[
\Psi (\rho _{1},\rho _{2})=\Psi _{1}(\rho _{1})+\Psi _{2}(\rho
_{2})
\]
and consequently,
\begin{equation}
\begin{array}{l}
p_l-p_v=\frac{1}{2}(\rho _{1v}-\rho _{1v_{o}})(\rho _{1l}+\rho
_{1l_{o}}-\rho _{1v}-\rho
_{1v_{o}})\Psi _{1}^{^{\prime \prime }}(\rho _{1v_{o}})\\
\quad \quad \quad \quad \quad  +\,\frac{1}{2}(\rho _{2v}-\rho
_{2v_{o}})(\rho _{2l}+\rho _{2l_{o}}-\rho _{2v}-\rho
_{2v_{o}})\Psi _{2}^{^{\prime \prime }}(\rho _{2v_{o}})
\end{array}
\end{equation}
where $p_{i}(\rho _{i})=\rho _{i}\Psi _{i}^{^{\prime }}(\rho
_{i})-\Psi _{i}(\rho _{i})+p_{i_{o}}$, with
  $p_{i_{o}}$ as an additive constant,
 denotes the partial pressure of the component $i$ of
the mixture in the bulk $v$ and $ p(\rho _{1},\rho _{2})=p_{1}(\rho
_{1})+p_{2}(\rho _{2})$. \\ Let us notice that $\rho _{1l}$ is the
liquid density when $\rho _{1v}$, $\rho _{2l}$, $\rho _{2v}$ denote
the gas densities. Moreover $\, \rho _{1v_{o}}\Psi _{1}^{^{\prime
\prime }}(\rho _{1v_{o}})= \displaystyle \frac{\partial
p_{1}}{\partial \rho _{1}}(\rho _{1v_{o}}) =  c_1^2 \, $ where $c_1$
is the isothermal celerity
 of the saturated vapor at the liquid-vapor equilibrium. With
 component
molar masses of the same order,  $c_1$ and the isothermal celerity of sound
in the gas are of the same magnitude. The second term of relation
(12) is negligible with respect to the first one and we obtain the
simple result
\begin{equation}
p_l-p_v\thickapprox (\rho _{1v}-\rho _{1v_{o}})\rho
_{1l_{o}}\Psi _{1}^{^{\prime \prime }}(\rho _{1v_{o}})
\end{equation}
Relation (10) yields
\begin{equation}
N=K\frac{\rho _{1m}-\rho _{1v}}{\rho _{1l}-\rho _{1v}}(\rho
_{1v}-\rho _{1v_{o}})^{3}\rho _{1l_{o}}^{3}\left[ \Psi _{1}^{^{\prime \prime }}(\rho _{1v_{o}})\right] ^{3}
\end{equation}
By noticing that $ \rho _{1l}-\rho _{1v}\thickapprox \rho
_{1l_{o}}$, one obtains
\[
N\thickapprox K(\rho _{1m}-\rho _{1v})(\rho _{1v}-\rho
_{1v_{o}})^{3}\rho _{1l_{o}}^{2}\left[ \Psi _{1}^{^{\prime \prime }}(\rho _{1v_{o}})\right] ^{3}
\]
expression which is maximum when  $\rho _{1v}=\displaystyle
\frac{3\, \rho _{1m}+\rho _{1v_0}}{4}$.\\
One deduces the maximum value of $N$
\begin{equation}
 N_{\max }=\frac{1}{32\pi \sigma
^{3}}(\rho _{1v}-\rho _{1v_{o}})^{4}\rho _{1l_{o}}^{2}\left[ \Psi
_{1}^{^{\prime \prime }}(\rho _{1v_{o}})\right] ^{3}
\end{equation}
Let us notice that $\displaystyle \Psi _{1}(\rho _{1v})-\Psi
_{1}(\rho _{1v_{o}})=\frac{1}{2}(\rho _{1v}-\rho
_{1v_{o}})^{2}\, \Psi _{1}^{^{\prime \prime }}(\rho
_{1v_{o}})$
 and by taking into account the choice of the
chemical potential components,\\ $\rho _{1v_{o}}\Psi _{1}^{\prime
}(\rho _{1v_{o}})-\rho _{1v}\Psi _{1}^{\prime }(\rho _{1v})=\rho
_{1v_{o}}\left[ \Psi _{1}^{\prime }(\rho _{1v_{o}})-\Psi
_{1}^{\prime }(\rho _{1v})\right] +(\rho _{1v_{o}}-\rho _{1v})\Psi
_{1}^{\prime }(\rho _{1v})=\rho _{1v_{o}}(\rho _{1v_{o}}-\rho
_{1v})\Psi _{1}^{^{\prime \prime }}(\rho _{1v_{o}})-(\rho
_{1v_{o}}- \rho _{1v})^{2}\Psi _{1}^{^{\prime \prime }}(\rho
_{1v_{o}})$. \\ Consequently, $p_{1v_{o}}-p_{1v}=\frac{1}{2}(\rho
_{1v_{o}}-\rho _{1v})(\rho _{1v_{o}}+\rho _{1v})\Psi
_{1}^{^{\prime \prime }}(\rho _{1v_{o}})\thickapprox \rho
_{1v_{o}}(\rho _{1v_{o}}-\rho _{1v})\Psi _{1}^{^{\prime \prime
}}(\rho _{1v_{o}})$, and relation (15) yields
\begin{equation}
N_{\max }=\frac{\rho _{1l_{o}}^{2}\left[ p_{1v}-p_{1v_{o}}\right] ^{4}}{%
32\pi \sigma ^{3}\rho _{1v_{o}}^{3}c_{1}^{2}}
\end{equation}
where $p_{1v}, p_{1v_{o}}$ denote the partial
pressures of the supersaturated vapor and the saturated vapor.\\
When $p_v =p_{1v}+p_{2}$ where $p_{2}$ denotes the partial pressure
of the gas  assumed to be independent of the quantity of vapor of
the component 1, one obtains
\begin{equation}
N_{\max }=\frac{\rho _{1l_{o}}^{2}\left[ p_v -p_{v_{o}}\right]
^{4}}{32\pi \sigma ^{3}\rho _{1v_{o}}^{3}c_{1}^{2}}
\end{equation}
This case corresponds to a vapor density smaller than the gas
density. It is generally the case for water vapor in air at
ambient temperature.\\
For example, we consider the physical values of liquid water at
the temperature of $25^\circ$ C: surface tension $\sigma = 72 $
dyne/cm, isothermal sound velocity $c_1 = 2.8\,\,\, 10^4$ cm/s,
$\rho_{1l_{o}}= 1 $ g/cm$^3$, $\rho_{1v_{o}}= 2.3\,\,\, 10^{-5}$
g/cm$^3$ and for $\displaystyle \frac {p_{1v}-
p_{1v_{o}}}{p_{1v_{o}}} = 5\,\,\, 10^{-5}$  corresponding to a
water vapor supersaturated of $0.2$ per $100$, one obtains a
maximum of droplets $N_{max}= 15 000 $ per cubic centimeter  of
gas \cite{crc}.

\section{ENERGY OF FORMATION OF DROPLETS}

From relations (7) and (9), the capillary energy of formation  of
$n$ droplets   is given by the expression
\[
E_c =\frac{3}{2}\frac{n}{K^{\frac{1}{3}}}\,
v_{l}^{\frac{2}{3}}\equiv \frac{3}{2}\frac{\rho _{1m}-\rho _{1v}}{\rho _{1l}-\rho
_{1v}}\, V\, (p_l-p_v)
\]
According to relation (10) one obtains,
\[
E_c
\thickapprox \frac{3}{ 2}\left( \rho _{1m}-\rho
_{1v}\right) \left( \rho _{1v}-\rho _{1v_{o}}\right) \Psi
_{1}^{^{\prime \prime }}(\rho_{1v_{o}})\, V
\]
From the choice of the mixture free energy, the total free energy of formation of the droplets
is $ E = E_0 + E_c$ where $E_0$
is the   bulk  free energy. One deduces
\[E_{o}=n\psi(l)v_{l}+\psi (v)v_{v}=\frac{V}{\rho _{1l}-\rho
_{1v}}\left[ \left( \rho _{1m}-\rho _{1v}\right) \psi(l)+(\rho
_{1l}-\rho _{1m})\psi (v) \right] \] By taking into account the
equality of the component chemical potentials  (11)  and due to the
fact that gas and  vapor are perfect gases, $\Psi _{\rho _{1}\rho
_{2}}^{^{\prime \prime }} = 0$, one obtains
\[
\begin{array}{l}
\displaystyle 2 \, (\rho _{1l}-\rho _{1v})\, e_{o}=\\
 \displaystyle (\rho _{1m}-\rho _{1v})
\left[ (\rho _{1l}-\rho _{1l_{o}})(\rho _{1v}-\rho _{1v_{o}})\Psi
_{1}^{^{\prime \prime }}(\rho_{1v_{o}})+ (\rho _{2l}-\rho _{2l_{o}})(\rho
_{2v}-\rho _{2v_{o}})\Psi _{2}^{^{\prime \prime }}(\rho_{2v_{o}})\right]\\
+ (\rho _{1l}-\rho _{1m}) \left[ (\rho _{1v}-\rho
_{1v_{o}})^{2}\Psi _{1}^{^{\prime \prime }}(\rho_{1v_{o}})+(\rho
_{2v}-\rho _{2v_{o}})^{2}\Psi _{2}^{^{\prime \prime
}}(\rho_{2v_{o}})\right]
\end{array}
\]
with
$\displaystyle e_{o}=\frac{E_{o}}{V}$. Due to the respective order
of $\rho_{1l}, \rho_{1v},\rho_{2l},\rho_{2v}$,  we deduce
 \[e_{o}\thickapprox \frac{1}{2}\left[(\rho
_{1v}-\rho _{1v_{o}})^{2}\Psi _{1}^{^{\prime \prime
}}(\rho_{1v_{o}})+(\rho _{2v}-\rho _{2v_{o}})^{2}\Psi _{2}^{^{\prime
\prime }}(\rho_{2v_{o}})\right]\]
 As indicated in paragraph 3,
 $\rho _{1vo}\Psi _{1}^{^{\prime \prime }}(\rho_{1v_{o}})\, $ and $\, \rho
_{2vo}\Psi _{2}^{^{\prime \prime }}(\rho_{2v_{o}})\, $ are of   same
order. Under usual thermodynamic conditions, the vapor density is
much lower than the  gas density; moreover $\rho _{2v}-\rho
_{2v_{o}}$ associated with a small variation of gas density is of the same order
 (or an order lower) than $ \rho
_{1v}-\rho _{1v_{o}}$ associated with variation of   vapor density. Then,
\[e_{o}\thickapprox \frac{1}{2}(\rho _{1v}-\rho _{1v_{o}})^{2}\Psi
_{1}^{^{\prime \prime }}(\rho_{1v_{o}})\] and
\[e \thickapprox \frac{1}{2} [(\rho _{1v}-\rho
_{1v_{o}})^{2}+3(\rho _{1m}-\rho _{1v})(\rho _{1v}-\rho
_{1v_{o}})] \Psi _{1}^{^{\prime \prime }}(\rho_{1v_{o}})\] is the
formation free energy of droplets per unit  volume.\\ This energy
has a maximum value when the number of drops per unit volume is
maximum. The value of this energy is
\[e_{\max }=(\rho _{1v}-\rho
_{1v_{o}})^{2}\Psi _{1}^{^{\prime \prime }}(\rho_{1v_{o}})\]
 and by a
 calculus as in paragraph 3, \[e_{\max }\thickapprox
\frac{\left[ p_{1v}-p_{1v_{o}}\right] ^{2}}{\rho
_{1vo}\, c_{1}^{2}}\] When $p_v =p_{1v}+p_{2}$, in the same way
that in paragraph 3,
\[e_{\max }\thickapprox \frac{\left[ p_v-p_{v_{o}}\right] ^{2}}{\rho
_{1vo}\, c_{1}^{2}}\] This value does not depend on the surface
tension.

\section { CONCLUSION}

The  approach  to determine the droplet number takes into account
the fact that the mixture free energy  is  a non-convex function of
the densities of  components. However, this result is not used in
explicit calculations and the maximum number of  droplets does not
take into account  the non-convex part of the  free energy since the
densities of  components are near equilibrium densities associated
with the liquid and its saturated vapor in presence of a gas. The
result is obtained for a gas and a vapor considered as perfect
gases. It cannot be extended to micro-bubbles  in a liquid because
the free energy of the liquid phase is not an additive function of
the mixture components.


\begin{thebibliography}{9}
\bibitem{Djikaev} Y.S. Djikaev and J. Teichmann,
J. Aerosol Sci.   30 (1999) 587.
\bibitem{Levdansky}  V.V. Levdansky et al,
Int. J. Heat and Mass Transfert.    45 (2002) 3831.
\bibitem{gouin} H. Gouin,   Eur. J. Mech./B Fluids.   9  (1990) 469.
\bibitem{Rocard}  Y. Rocard, Thermodynamique, Masson,
Paris, 1952.
\bibitem{Rowlinson} J.S. Rowlinson and B. Widom, Molecular theory of capillarity,
Clarendon Press, Oxford,  1984.
\bibitem{crc}R.C. Weast, Handbook of Chemistry and Physics, 65th edition, CRC
Press, Boca Raton, Florida, 1992.
\end{thebibliography}
\end{document}